\documentclass[reviewcopy]{elsart}
\usepackage{amsmath}
\usepackage{graphicx}
\usepackage{cite}
\usepackage{lineno}

\journal{Surface Science}

\begin{document}
\begin{frontmatter}
  \title{Newns-Anderson model of chemicurrents in H/Cu and H/Ag}
  
  \author[bath]{M. S. Mizielinski\corauthref{cor}},
  \corauth[cor]{Corresponding author}
  \ead{M.S.Mizielinski@bath.ac.uk}
  \author[bath]{D. M. Bird},
  \address[bath]{Department of Physics, University of Bath, Bath, BA2 7AY, UK}
  \author[liv]{M. Persson} and
  \author[liv]{S. Holloway}
  \address[liv]{Surface Science Research Centre, University of Liverpool,
    Liverpool, L69 3BX, UK}

  \begin{abstract}
    
    The excitation of the electronic system induced by the adsorption of
    a hydrogen atom on the (111) surfaces of copper and silver is investigated
    using the time-dependent, mean-field Newns-Anderson model.
    Parameters for the model are obtained by fitting to density functional
    theory calculations, allowing the charge and energy transfer between
    adsorbate and surface to be calculated, together with the spectrum of
    electronic excitations. 
    These results are used to make direct comparisons with experimental
    measurements of chemicurrents, yielding good agreement for both the
    magnitude of the current and the ratio of the currents for H and D
    adsorption. 
    
  \end{abstract}

\end{frontmatter}

\newcommand{\ep}{\epsilon}
\newcommand{\ea}{\epsilon_a}
\newcommand{\ebar}{\bar{\epsilon}_{a\sigma}}
\newcommand{\ebarad}{\bar{\epsilon}_{a\sigma}^{(ad)}}
\newcommand{\nasad}{n_{a\sigma}^{(ad)}}
\newcommand{\nas}{n_{a\sigma}}
\newcommand{\erfc}{\textrm{erfc}}
\newcommand{\nsex}{n^{(ex)}_\sigma}
\newcommand{\nex}{n^{(ex)}}
\newcommand{\Teff}{T^{\textrm{(eff)}}}
\newcommand{\Tchemi}{T^{\textrm{(chemi)}}}
\newcommand{\Pchemie}{P^{\textrm{(chemi)}}_e}
\newcommand{\Pchemih}{P^{\textrm{(chemi)}}_h}
\newcommand{\Pchemi}{P^{\textrm{(chemi)}}}

\section{Introduction}
\label{sec:intro}
\linenumbers

Until recently the direct observation of non-adiabatic dissipation
of energy into electronic excitations during adsorption events has been
limited to highly energetic processes, such as the oxidation of alkali and
alkali-earth metals \cite{greber97}. 
Such reactions can result in chemiluminescence or the ejection of
exo-electrons. 
Investigation of low-energy electronic excitations has been restricted 
by the difficulties involved in making experimental measurements.
However, two recent series of experiments have provided the first direct
observations of the excitation of relatively low-energy electrons and holes. 
White, Wodtke and co-workers \cite{white05,white06,wodtke08} observed
exo-electron excitation when a low-workfunction, caesium-doped gold surface
was exposed to a beam of vibrationally excited NO molecules.
They suggested that during the vibration of the NO molecules the ground
electronic state oscillates between the neutral and negative ion and rapid
transfer of an electron between the surface and the molecule during this
oscillation leads to the excitation of the electronic system \cite{white06}.

Here we are interested in the chemicurrent experiments performed 
by Nienhaus and coworkers
\cite{nienhaus99,gergen01,nienhaus02a,nienhaus02b,krix07}.
These experiments involve the fabrication of Schottky diodes,
consisting of a thin, $\sim$100 \AA{}, metal film deposited onto a doped
silicon wafer with electrical contacts made to the film and the back of the
wafer. 
On exposure to beams of atomic hydrogen, hot electrons or holes with
sufficient energy to traverse the metal film and cross the Schottky barrier at
the metal-semiconductor interface were measured as a
chemically-induced-reverse-current or `chemicurrent'.
These devices have been used to investigate differences in the adsorption of
hydrogen isotopes \cite{nienhaus99,krix07} as well as a range of other
adsorbates \cite{gergen01,nienhaus02b}. 
Similar devices have also been used to study other surface phenomena
\cite{hasselbrink07} including adsorption and desorption in H/Au
\cite{mildner06} and chemiluminescence in O/Mg \cite{nienhaus06}.  

Theoretical modelling of the electronic excitations generated during
the adsorption of hydrogen atoms on metal surfaces has been performed recently
using three techniques: electronic friction based methods
\cite{trail01,persson04, luntz05, trail02,trail03}, 
the time-dependent, mean-field Newns-Anderson model
\cite{anderson61,newns69,bird04,mizielinski05,mizielinski07,thesis,bird08} and
time-dependent density functional theory (TDDFT) 
\cite{lindenblatt06nm,lindenblatt06ss,lindenblatt06}. 

Electronic friction methods use a nearly-adiabatic approximation in which the
time-dependent perturbation of the electronic system is assumed to be weak and
slow.
This leads to a description of energy transfer equivalent to that induced by a
simple frictional force.
The electronic friction coefficient can be calculated through ab-initio
methods\cite{trail01}, and has been widely used in the study of surface
dynamics, including the damping of vibrations in adsorbed
molecules\cite{persson04} and desorption dynamics \cite{luntz05}.
The energy distribution of excited electron-hole pairs can be obtained
by coupling this friction description to the forced oscillator model
\cite{trail02,trail03}.

However, electronic friction calculations exhibit problematic features when
considering strongly non-adiabatic behaviour. 
Trail and coworkers \cite{trail02,trail03} found that ab-initio calculations
of the friction coefficient for an H-atom above a copper (111) surface yield a
singularity at an altitude of 2.4 \AA{} above the atop site.
This unphysical feature was linked to the change in the ground
state from being spin-polarised (H-atom far from the surface) to unpolarised
(H-atom close to the surface). 
To avoid this problem a somewhat arbitrary choice was made to constrain the
spin of the DFT calculations to be constant, giving a non-singular variation
for the friction coefficient.
Calculations using these constrained results yielded probabilities for
electrons being excited over a Schottky barrier which were in line with the
experimental results of Nienhaus and coworkers.
The use of a spin-constrained calculation is not, however, a satisfactory
solution to the problem, and other methods have been sought which can describe
systems which experience a spin-transition. 

The time-dependent, mean-field Newns-Anderson model \cite{anderson61,newns69},
used in our previous work \cite{bird04,mizielinski05,mizielinski07,thesis},
provides a straightforward way to study the spin-transition in a fully
non-adiabatic fashion. 
This model describes the interaction of a single adsorbate orbital, containing
a pair of coupled energy levels, with a broad band of metal states.
Time-dependence is included through the movement of the adsorbate energy
levels relative to the Fermi level and the variation of the adsorbate-metal 
interaction.  
Within the mean-field and wide-band approximations expressions describing the
time-evolution of the adsorbate energy level occupations, the non-adiabatic
transfer of energy to the surface \cite{mizielinski05,thesis}, and the
spectrum of electronic excitations have been derived \cite{mizielinski07,thesis}.

TDDFT calculations take the set of Kohn-Sham wavefunctions for a given system,
generated from a conventional static DFT calculation, and evolve them through
the time-dependent Schr\"odinger equation, using Ehrenfest dynamics for the
nuclear motion. 
This technique has been used by Lindenblatt and Pehlke
\cite{lindenblatt06nm,lindenblatt06ss,lindenblatt06}
to investigate the interaction of hydrogen atoms with an aluminium surface,
yielding results for the non-adiabatic energy transfer and the spectrum of
electronic excitations.
However, computational constraints restrict the application of this technique
to consideration of light elements only, and the restricted basis set leads to
somewhat noisy results.
A recent comparison of TDDFT and Newns-Anderson results for the H/Al system
\cite{bird08} has shown good agreement in these two descriptions of
non-adiabatic behaviour.

In our previous publications we have introduced and demonstrated the
properties of the time-dependent, mean-field Newns-Anderson model  
\cite{mizielinski05,mizielinski07} using simple parameter variations to
explore the non-adiabatic evolution of the adsorbate-metal system.
Here, we use this model to analyse systems of direct relevance
to the chemicurrent experiments described above: hydrogen and deuterium atoms
approaching the copper and silver surfaces.
This work is presented in two steps.
In section \ref{sec:params} the method used to generate appropriate parameters
for the H/Cu and H/Ag systems is described.
These parameters are then used in section \ref{sec:results} to make
comparisons between theoretical and experimental results for both the size of
the chemicurrent and isotopic ratios on each metal surface.
Conclusions are drawn from these results in section \ref{sec:conclusions}.

\section{Parameterisation of the Newns-Anderson model}
\label{sec:params}

Within the wide-band and mean-field approximations the Newns-Anderson model
can be parameterised through the position of the bare adsorbate level $\ea$,
the width of the adsorbate resonance $\Gamma$ and the intra-adsorbate Coulomb
repulsion energy $U$ \cite{anderson61,mizielinski05,mizielinski07,thesis}. 
Values of these parameters, as a function of H-atom altitude, are found by
fitting the adiabatic solution of the mean-field Newns-Anderson model to
ground-state DFT calculations.
The time-dependence is then obtained by choosing a trajectory which links
altitude to time in a realistic way.

A series of DFT calculations of hydrogen atoms above the copper and silver
(111) surfaces have been performed using the CASTEP \cite{castep} code.
The surfaces were modelled with a slab geometry consisting of five layers of
atoms and an equivalent vacuum gap above, with lattice parameters fixed at the
experimental bulk values of 3.614 \AA{} and 4.085 
\AA{} for copper and silver respectively \cite{crc}. 
The hydrogen atom was placed at a set of altitudes above an atop site of the
surface. 
Both systems were represented using a 2 $\times$ 2 in-plane supercell, and
ultra-soft pseudopotentials were used for both the metal and the hydrogen
atoms. 
The Perdew-Burke-Ernzerhof (PBE) exchange-correlation functional was used, and
a Fermi surface smearing of 0.25 eV was applied. 
Sampling of the surface Brillouin zone was performed using 54 k-points, and
plane-wave cutoffs of 290 eV and 300 eV were used for the copper and silver
surfaces respectively. 

\begin{figure}[p]
  \begin{center}
  \begin{picture}(0,0)%
    \includegraphics{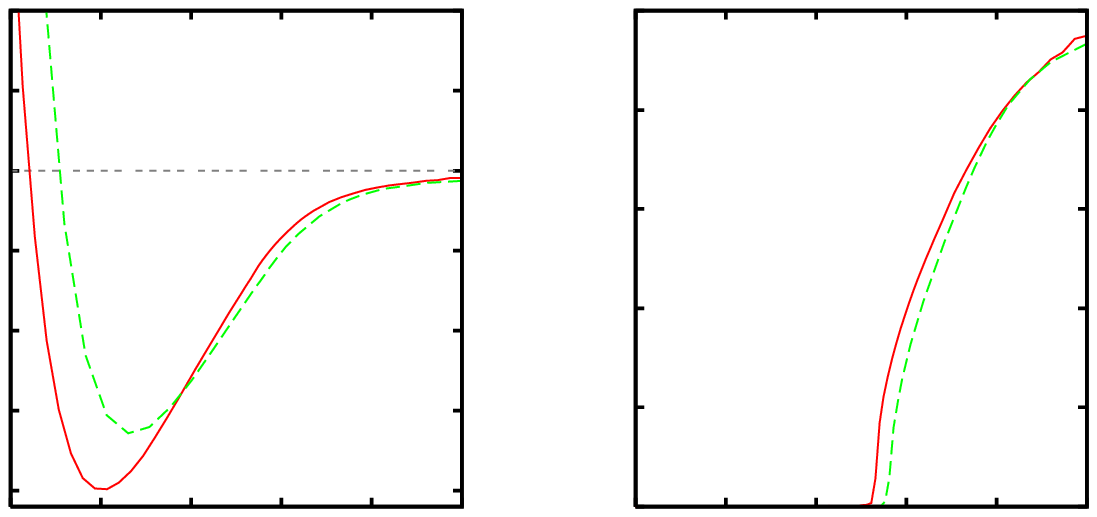}%
  \end{picture}%
  \begingroup
  \setlength{\unitlength}{0.1bp}%
  \begin{picture}(3600,1700)(0,0)%
    \put(3450, 100){\makebox(0,0){ 3.5}}%
    \put(3190, 100){\makebox(0,0){ 3.0}}%
    \put(2930, 100){\makebox(0,0){ 2.5}}%
    \put(2670, 100){\makebox(0,0){ 2.0}}%
    \put(2410, 100){\makebox(0,0){ 1.5}}%
    \put(2150, 100){\makebox(0,0){ 1.0}}%
    \put(2800, -100){\makebox(0,0)[c]{H-atom altitude (\AA{})}}%
     
    \put(2100,1628){\makebox(0,0)[r]{ 1.0}}%
    \put(2100,1342){\makebox(0,0)[r]{ 0.8}}%
    \put(2100,1057){\makebox(0,0)[r]{ 0.6}}%
    \put(2100, 771){\makebox(0,0)[r]{ 0.4}}%
    \put(2100, 486){\makebox(0,0)[r]{ 0.2}}%
    \put(2100, 200){\makebox(0,0)[r]{ 0.0}}%
    \put(1800, 914){\rotatebox{90}{\makebox(0,0)[c]{Spin polarisation}}}

    \put(1650, 100){\makebox(0,0){ 3.5}}%
    \put(1390, 100){\makebox(0,0){ 3.0}}%
    \put(1130, 100){\makebox(0,0){ 2.5}}%
    \put(870 , 100){\makebox(0,0){ 2.0}}%
    \put(610 , 100){\makebox(0,0){ 1.5}}%
    \put(350 , 100){\makebox(0,0){ 1.0}}%
    \put(1000, -100){\makebox(0,0)[c]{H-atom altitude (\AA{})}}%

    \put(300 ,1628){\makebox(0,0)[r]{ 1.0}}%
    \put(300 ,1398){\makebox(0,0)[r]{ 0.5}}%
    \put(300 ,1167){\makebox(0,0)[r]{ 0.0}}%
    \put(300 , 937){\makebox(0,0)[r]{$-$0.5}}%
    \put(300 , 707){\makebox(0,0)[r]{$-$1.0}}%
    \put(300 , 476){\makebox(0,0)[r]{$-$1.5}}%
    \put(300 , 246){\makebox(0,0)[r]{$-$2.0}}%
    \put(-100,  914){\rotatebox{90}{\makebox(0,0)[c]{Potential energy (eV)}}}

    \put(1550,1500){\makebox(0,0)[c]{(a)}}
    \put(2250,1500){\makebox(0,0)[c]{(b)}}

  \end{picture}%

  \endgroup
  \end{center}
  \vspace{0.5cm}
  \caption{(a) Surface potential well above the atop site and (b) spin
    polarisation of the H/Cu (solid-red lines) and H/Ag (dashed green lines)
    systems.} 
    \label{fig:pes}
\end{figure}

\begin{figure}[p]
  \begin{center}
  \begin{picture}(0,0)%
    \includegraphics{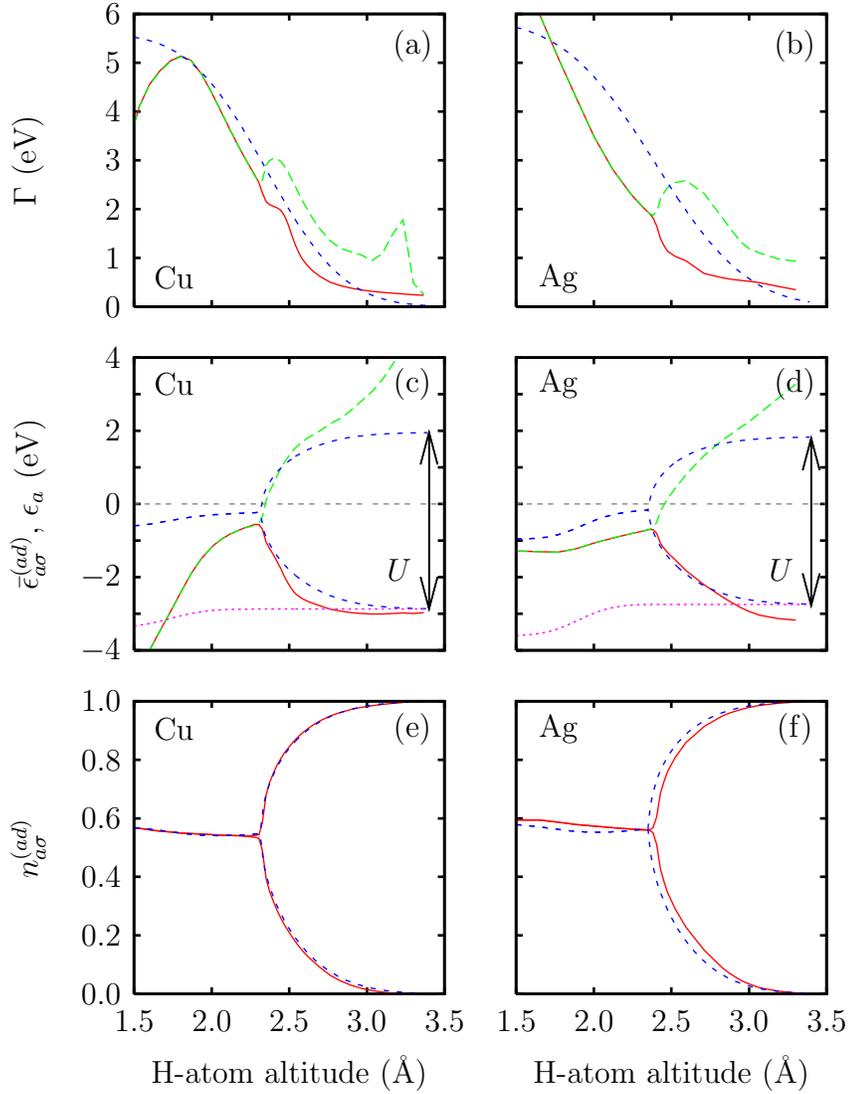}%
  \end{picture}%
  \begingroup
  \setlength{\unitlength}{0.1bp}%
  \begin{picture}(3000,3900)(0,0)%
    \put(2910,100){\makebox(0,0){3.5}}%
    \put(2618,100){\makebox(0,0){3.0}}%
    \put(2325,100){\makebox(0,0){2.5}}%
    \put(2033,100){\makebox(0,0){2.0}}%
    \put(1740,100){\makebox(0,0){1.5}}%
    \put(2325, -100){\makebox(0,0)[c]{H-atom altitude (\AA{})}}%
    
    \put(1470, 100){\makebox(0,0){3.5}}%
    \put(1178, 100){\makebox(0,0){3.0}}%
    \put(885 , 100){\makebox(0,0){2.5}}%
    \put(593 , 100){\makebox(0,0){2.0}}%
    \put(300 , 100){\makebox(0,0){1.5}}%
    \put(885, -100){\makebox(0,0)[c]{H-atom altitude (\AA{})}}%

    \put(250 ,1304){\makebox(0,0)[r]{1.0}}%
    \put(250 ,1083){\makebox(0,0)[r]{0.8}}%
    \put(250 , 862){\makebox(0,0)[r]{0.6}}%
    \put(250 , 642){\makebox(0,0)[r]{0.4}}%
    \put(250 , 421){\makebox(0,0)[r]{0.2}}%
    \put(250 , 200){\makebox(0,0)[r]{0.0}}%
    \put(-100, 752){\rotatebox{90}{\makebox(0,0)[c]{$\nasad$}}}

    \put(250 ,2600){\makebox(0,0)[r]{4}}%
    \put(250 ,2324){\makebox(0,0)[r]{2}}%
    \put(250 ,2048){\makebox(0,0)[r]{0}}%
    \put(250 ,1772){\makebox(0,0)[r]{$-$2}}%
    \put(250 ,1496){\makebox(0,0)[r]{$-$4}}%
    \put(-100,2048){\rotatebox{90}{\makebox(0,0)[c]{$\ebarad$, $\ea$ (eV)}}}

    \put(250 ,3896){\makebox(0,0)[r]{6}}%
    \put(250 ,3712){\makebox(0,0)[r]{5}}%
    \put(250 ,3528){\makebox(0,0)[r]{4}}%
    \put(250 ,3344){\makebox(0,0)[r]{3}}%
    \put(250 ,3160){\makebox(0,0)[r]{2}}%
    \put(250 ,2976){\makebox(0,0)[r]{1}}%
    \put(250 ,2792){\makebox(0,0)[r]{0}}%
    \put(-100,3344){\rotatebox{90}{\makebox(0,0)[c]{$\Gamma$ (eV)}}}

    \put(1350,3775){\makebox(0,0)[c]{(a)}}
    \put(2790,3775){\makebox(0,0)[c]{(b)}}
    \put(1350,2500){\makebox(0,0)[c]{(c)}}
    \put(2790,2500){\makebox(0,0)[c]{(d)}}
    \put(1350,1200){\makebox(0,0)[c]{(e)}}    
    \put(2790,1200){\makebox(0,0)[c]{(f)}}

    \put( 450,2900){\makebox(0,0)[c]{Cu}}
    \put( 450,2500){\makebox(0,0)[c]{Cu}}
    \put( 450,1200){\makebox(0,0)[c]{Cu}}
    \put(1900,2900){\makebox(0,0)[c]{Ag}}
    \put(1900,2500){\makebox(0,0)[c]{Ag}}    
    \put(1900,1200){\makebox(0,0)[c]{Ag}}

    \put(1300,1800){\makebox(0,0)[c]{$U$}}
    \put(2740,1800){\makebox(0,0)[c]{$U$}}
  \end{picture}%

  \endgroup
  \end{center}
  \vspace{0.5cm}
  \caption{Parameter variations for the H/Cu [(a), (c) and (e)] and H/Ag [(b), 
    (d) and (f)] systems. In (a) and (b) solid red and long-dashed green lines 
    denote the fitted resonance widths for majority and minority spin 
    respectively, while the medium-dashed blue line denotes the error function
    fit used in later calculations. (c) and (d) show the energy levels
    with solid red (majority) and long-dashed green (minority) lines being
    fits to the DFT PDOS, and the medium-dashed blue and short-dashed magenta
    lines denoting the effective adsorbate energy level $\ebarad$ and the
    error function fit to $\ea$ respectively.  Arrows in (c) and (d) indicate
    the value of $U$.  The bottom two panels, (e) and (f), show the
    occupations of the adsorbate levels arising from the DFT calculations
    (solid red lines) and from the chosen parameter variations (medium-dashed
    blue lines).}
    \label{fig:params}
\end{figure}

A total of 88 calculations were performed for the two systems with H-atom
altitudes varied between 1 and 3.5 \AA{}.
The potential energy curves and the spin-polarisation are plotted in
Fig. \ref{fig:pes}.
Panel (b) shows the square-root like, second-order transition in the
spin-polarisation of the hydrogen-metal systems, a characteristic of 
mean-field theories.
At each altitude the projected density of states (PDOS) onto the hydrogen 1s
orbital is calculated for fitting to the Newns-Anderson model.
In the adiabatic limit of the wide-band, mean-field Newns-Anderson model the
two adsorbate resonances, one for each spin $\sigma$, are Lorentzian in shape
with width $\Gamma$, centred on the effective energy levels $\ebarad$
\cite{mizielinski05}. 
From the PDOS, fitted values for $\Gamma$ and $\ebarad$ have been extracted
and are plotted in Figs \ref{fig:params}(a)-(d).
While these results could be used directly to obtain a variation for the bare
energy level $\ea$ and the value of $U$, it is important to consider whether
this would provide the best description of the excitation process in the H/Cu
and H/Ag systems.
In previous work \cite{mizielinski07,thesis} we have shown that the transfer
of charge between the adsorbate and surface can significantly alter the
excitation spectrum.
An excess of high-energy electrons is produced if there is a net electron
transfer to the surface, while an excess of high-energy holes is generated if
adsorption is accompanied by electron transfer to the adsorbate.
It is therefore important that the occupations of the adsorbate level  are
correctly represented in order to obtain reliable results. 

We have therefore devised a procedure which gives parameterisations of
$\Gamma$ and $\ea$, and the value of $U$, which are consistent with a desired
variation of the adsorbate level occupation. 
The target occupations are calculated by integrating the DFT generated H-atom
PDOS up to the Fermi level, and are shown in Figs \ref{fig:params}(e) and (f).
Error functions are then used to fit the variation of the resonance width and
bare energy level with altitude, with the constraint that the adsorbate
occupations are consistent with those obtained from the DFT PDOS. 
The procedure for obtaining these parameters is described in detail in
Ref. \cite{bird08}, and yields the following results for the H/Cu and H/Ag systems;

\begin{subequations}
\begin{eqnarray}
  \textrm{H/Cu}:\hspace{0.5cm} 
  \frac{\ea}{\textrm{eV}} &=&
  -2.872-0.263\ \erfc\left(3.717\left(\frac{s}{\textrm{\AA{}}}-1.729\right)\right), \\
  \label{eq:HCu_ea}
  \frac{\Gamma}{\textrm{eV}} &=&
  -8.020\times10^{-5} +
  2.805\ \erfc\left(1.796\left(\frac{s}{\textrm{\AA{}}}-2.352\right)\right),\\
  \label{eq:HCu_gamma}
  \frac{U}{\textrm{eV}} &=& 4.827,
  \label{eq:HCu_U}
\end{eqnarray}
\label{eq:HCu}
\end{subequations}
\begin{subequations}
\begin{eqnarray}
  \textrm{H/Ag}:\hspace{0.5cm} 
  \frac{\ea}{\textrm{eV}} &=&
  -2.774
  -0.430\ \erfc\left(4.076\left(\frac{s}{\textrm{\AA{}}}-1.901\right)\right), \\
  \label{eq:HAg_ea}
  \frac{\Gamma}{\textrm{eV}} &=&
  -1.748\times10^{-3} +
  2.944\ \erfc\left(1.514\left(\frac{s}{\textrm{\AA{}}}-2.396\right)\right),\\
  \label{eq:HAg_gamma}
  \frac{U}{\textrm{eV}} &=& 4.574,
  \label{eq:HAg_U}
\end{eqnarray}
\label{eq:HAg}
\end{subequations}

where $s$ is the altitude of the H-atom above the atop site.
These variations are plotted, along with the fits to the DFT PDOS in Figs
\ref{fig:params}(a)-(d), with the resulting adsorbate level occupations in
panels (e) and (f).

To complete the parameterisation of the Newns-Anderson model we require the
variation of $\Gamma$ and $\ea$ with time.
We have chosen to use a constant total energy trajectory with an initial
kinetic energy at 4 \AA{} of 25 meV, where the H-atom is accelerated in the
potential energy curves shown in Fig \ref{fig:pes}.
As we are particularly interested in the effects of the spin-transition,
calculations are terminated when the adsorbate reaches the back of the surface
potential well, i.e when the adsorbate reaches 1.1 \AA{} or 1.25 \AA{} for the
copper and silver surfaces respectively.
Isotope effects (explored experimentally by Krix, Nienhaus and co-workers
\cite{krix07}) can be investigated by changing the adsorbate mass in 
the trajectory calculations.

\section{Results}
\label{sec:results}

In this section we use the parameters derived above and the computational
model described previously \cite{mizielinski05,mizielinski07,thesis} to
investigate the non-adiabatic behaviour of the H/Cu and H/Ag systems. 
In addition to $\Gamma(t)$, $\ea(t)$ and $U$, the
computation of the adsorbate occupations, energy transfer rates and
excitation spectra requires a set of energy grids for the evaluation of
integrals. 
Here, a 128,001 point energy grid covering the range $-$90 to 10 eV relative
to the Fermi level has been used with numerical methods equivalent to those
discussed previously \cite{mizielinski07,thesis}.
A system temperature of 175 K is used in all calculations.
 
\begin{figure}[p]
  \begin{center}
  \begin{picture}(0,0)%
    \includegraphics{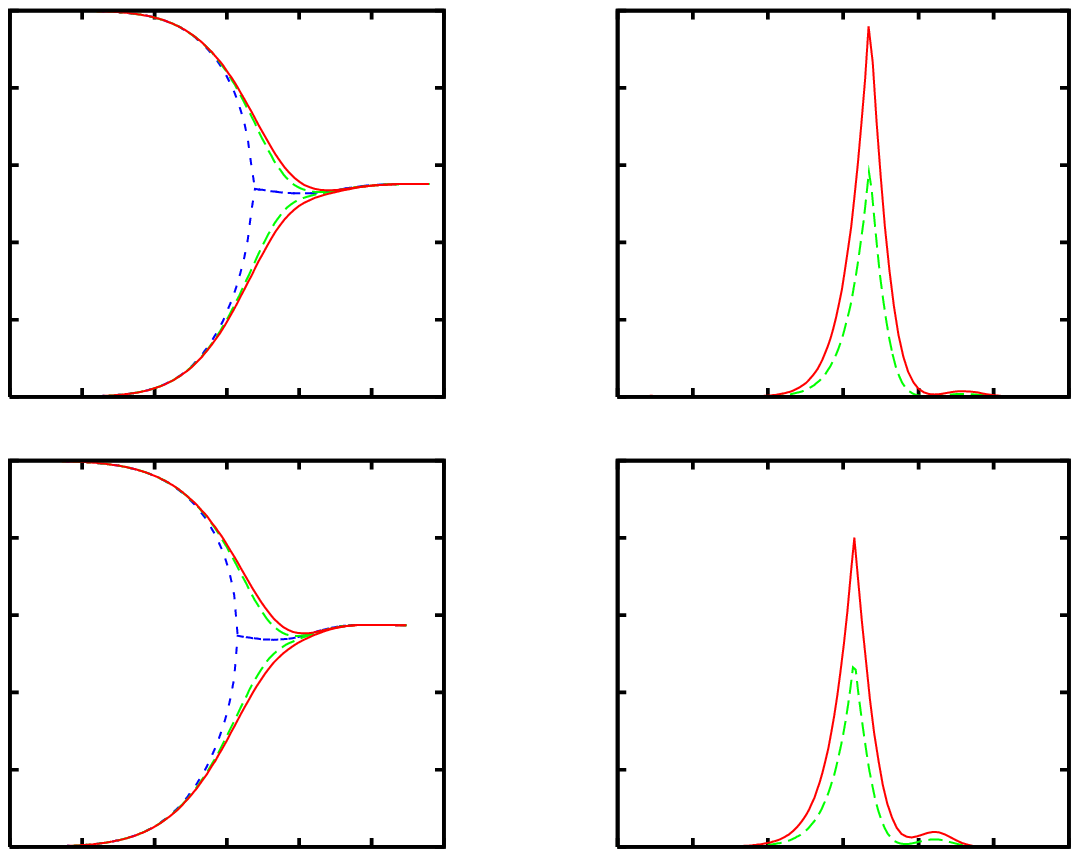}%
  \end{picture}%
  \begingroup
  \setlength{\unitlength}{0.1bp}%
  \begin{picture}(3600,2800)(0,0)%
    \put(2800,   0){\makebox(0,0){Adsorbate altitude (\AA{})}}%
    \put(1800, 856){\rotatebox{90}{\makebox(0,0)[c]{$\dot{E}_{non-ad}$ (meV/fs)}}}%

    \put(2150, 200){\makebox(0,0)[c]{4}}%
    \put(2583, 200){\makebox(0,0)[c]{3}}%
    \put(3017, 200){\makebox(0,0)[c]{2}}%
    \put(3450, 200){\makebox(0,0)[c]{1}}%

    \put(2100,1412){\makebox(0,0)[r]{50}}%
    \put(2100,1190){\makebox(0,0)[r]{40}}%
    \put(2100, 967){\makebox(0,0)[r]{30}}%
    \put(2100, 745){\makebox(0,0)[r]{20}}%
    \put(2100, 522){\makebox(0,0)[r]{10}}%
    \put(2100, 300){\makebox(0,0)[r]{ 0}}%
    
    \put(1025,   0){\makebox(0,0){Adsorbate altitude (\AA{})}}%
    \put(0   , 856){\rotatebox{90}{\makebox(0,0)[c]{$\nas$, $\nasad$}}}

    \put(400 , 200){\makebox(0,0)[c]{4}}%
    \put(817 , 200){\makebox(0,0)[c]{3}}%
    \put(1233, 200){\makebox(0,0)[c]{2}}%
    \put(1650, 200){\makebox(0,0)[c]{1}}%
    \put(350 ,1412){\makebox(0,0)[r]{1.0}}%
    \put(350 ,1190){\makebox(0,0)[r]{0.8}}%
    \put(350 , 967){\makebox(0,0)[r]{0.6}}%
    \put(350 , 745){\makebox(0,0)[r]{0.4}}%
    \put(350 , 522){\makebox(0,0)[r]{0.2}}%
    \put(350 , 300){\makebox(0,0)[r]{0.0}}%
    
    \put(1800,2152){\rotatebox{90}{\makebox(0,0)[c]{$\dot{E}_{non-ad}$ (meV/fs)}}}%
    \put(2100,2708){\makebox(0,0)[r]{50}}%
    \put(2100,2486){\makebox(0,0)[r]{40}}%
    \put(2100,2263){\makebox(0,0)[r]{30}}%
    \put(2100,2041){\makebox(0,0)[r]{20}}%
    \put(2100,1818){\makebox(0,0)[r]{10}}%
    \put(2100,1596){\makebox(0,0)[r]{ 0}}%

    \put(0   ,2152){\rotatebox{90}{\makebox(0,0)[c]{$\nas$, $\nasad$}}}%

    \put(350 ,2708){\makebox(0,0)[r]{1.0}}%
    \put(350 ,2486){\makebox(0,0)[r]{0.8}}%
    \put(350 ,2263){\makebox(0,0)[r]{0.6}}%
    \put(350 ,2041){\makebox(0,0)[r]{0.4}}%
    \put(350 ,1818){\makebox(0,0)[r]{0.2}}%
    \put(350 ,1596){\makebox(0,0)[r]{0.0}}%
    
    \put(1550,2600){\makebox(0,0)[c]{(a)}}
    \put(3350,2600){\makebox(0,0)[c]{(b)}}
    \put(1550,1300){\makebox(0,0)[c]{(c)}}
    \put(3350,1300){\makebox(0,0)[c]{(d)}}

    \put( 550,2600){\makebox(0,0)[c]{Cu}}
    \put(2300,2600){\makebox(0,0)[c]{Cu}}
    \put( 550,1300){\makebox(0,0)[c]{Ag}}
    \put(2300,1300){\makebox(0,0)[c]{Ag}}
    
  \end{picture}%

  \endgroup
  \end{center}
  \vspace{0.5cm}
  \caption{Adsorbate level occupations, (a) and (c), and energy transfer rate,
    (b) and (d) as a function of altitude for hydrogen (solid red lines) and
    deuterium atoms (long-dashed green lines) approaching the copper, (a) and
    (b), and silver, (c) and (d), surfaces.  Medium-dashed blue lines in
    panels (a) and (c) denote the adiabatic occupations for the two systems.}
    \label{fig:n_E}
\end{figure}

Fig. \ref{fig:n_E} shows the time-evolving charge and energy transfer
behaviour for hydrogen and deuterium atoms approaching the copper and silver
(111) surfaces.
As the adsorbates approach the metal surfaces the adiabatic occupations for
the majority and minority spins converge on one another resulting in a sharp
spin transition at 2.3-2.4 \AA{}.
The time-dependent occupations overshoot this spin-transition, with smaller
differences $\nas-\nasad$ for the slower deuterium atoms in comparison with
those for hydrogen. 
Associated with this overshoot of the adiabatic spin-transition is a
non-adiabatic transfer of energy to the metal surface, the rate of which is
shown in panels (b) and (d) of Fig. \ref{fig:n_E}.
Each system shows a sharp peak in this energy transfer rate at the spin
transition  with a small secondary peak (most prominent for the silver
surface) just below 2 \AA{}.
This secondary peak is driven by the variation of the bare adsorbate level
$\ea$, (see Figs. \ref{fig:params}(c) and (d)), while the large increase in
$\Gamma$ is responsible for most of the non-adiabatic behaviour close to the
spin-transition. 

By integrating over each trajectory the energy dissipated into electron-hole
pairs during the approach to the surface can be obtained. 
We find that H (D) atoms approaching the copper surface deposit 115 meV (88
meV), and those approaching the silver surface deposit 105 meV (80 meV).
It is important to note that these energy transfers are expectation values and
as such are an average over many trajectories.
It is therefore possible for an electron hole-pair to have more energy than
this average energy transfer, but with a limited probability.

\begin{figure}[p]
  \begin{center}
  \begin{picture}(0,0)%
    \includegraphics{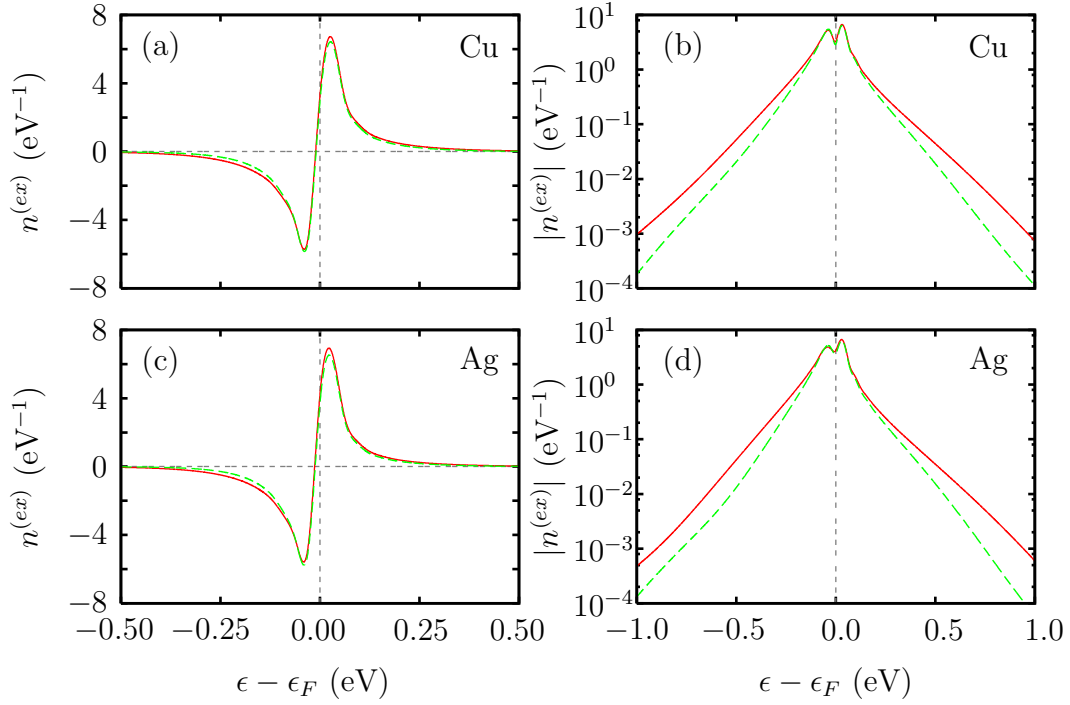}%
  \end{picture}%
  \begingroup
  \setlength{\unitlength}{0.1bp}%
  \begin{picture}(3500,2200)(0,0)%

    \put(3610,   0){\makebox(0,0){   1.0}}%
    \put(3232,   0){\makebox(0,0){   0.5}}%
    \put(2845,   0){\makebox(0,0){   0.0}}%
    \put(2458,   0){\makebox(0,0){$-$0.5}}%
    \put(2095,   0){\makebox(0,0){$-$1.0}}%
    \put(2845,-200){\makebox(0,0)[c]{$\ep-\ep_F$ (eV)}}%

    \put(1625,   0){\makebox(0,0)[c]{   0.50}}%
    \put(1262,   0){\makebox(0,0)[c]{   0.25}}%
    \put(875 ,   0){\makebox(0,0)[c]{   0.00}}%
    \put(488 ,   0){\makebox(0,0)[c]{$-$0.25}}%
    \put(125 ,   0){\makebox(0,0)[c]{$-$0.50}}%
    \put(875 ,-200){\makebox(0,0)[c]{$\ep-\ep_F$ (eV)}}%

    \put(2085,1131){\makebox(0,0)[r]{ 10$^{1\phantom{-}}$}}%
    \put(2085, 925){\makebox(0,0)[r]{ 10$^{0\phantom{-}}$}}%
    \put(2085, 719){\makebox(0,0)[r]{ 10$^{-1}$}}%
    \put(2085, 513){\makebox(0,0)[r]{ 10$^{-2}$}}%
    \put(2085, 307){\makebox(0,0)[r]{ 10$^{-3}$}}%
    \put(2085, 100){\makebox(0,0)[r]{ 10$^{-4}$}}%
    \put(1760, 615){\rotatebox{90}{\makebox(0,0)[c]{$\vert\nex\vert$ (eV$^{-1}$)}}}

    \put(2085,2319){\makebox(0,0)[r]{ 10$^{1\phantom{-}}$}}%
    \put(2085,2113){\makebox(0,0)[r]{ 10$^{0\phantom{-}}$}}%
    \put(2085,1907){\makebox(0,0)[r]{ 10$^{-1}$}}%
    \put(2085,1701){\makebox(0,0)[r]{ 10$^{-2}$}}%
    \put(2085,1495){\makebox(0,0)[r]{ 10$^{-3}$}}%
    \put(2085,1288){\makebox(0,0)[r]{ 10$^{-4}$}}%
    \put(1760,1803){\rotatebox{90}{\makebox(0,0)[c]{$\vert\nex\vert$ (eV$^{-1}$)}}}

    \put(100 ,2319){\makebox(0,0)[r]{   8}}%
    \put(100 ,2061){\makebox(0,0)[r]{   4}}%
    \put(100 ,1803){\makebox(0,0)[r]{   0}}%
    \put(100 ,1545){\makebox(0,0)[r]{$-$4}}%
    \put(100 ,1288){\makebox(0,0)[r]{$-$8}}%
    \put(-200,1803){\rotatebox{90}{\makebox(0,0)[c]{$\nex$ (eV$^{-1}$)}}}

    \put(100 ,1131){\makebox(0,0)[r]{   8}}%
    \put(100 , 873){\makebox(0,0)[r]{   4}}%
    \put(100 , 615){\makebox(0,0)[r]{   0}}%
    \put(100 , 357){\makebox(0,0)[r]{$-$4}}%
    \put(100 , 100){\makebox(0,0)[r]{$-$8}}%
    \put(-200, 615){\rotatebox{90}{\makebox(0,0)[c]{$\nex$ (eV$^{-1}$)}}}
    
    \put(300 ,2200){\makebox(0,0)[c]{(a)}}
    \put(2270,2200){\makebox(0,0)[c]{(b)}}
    \put(300 ,1012){\makebox(0,0)[c]{(c)}}
    \put(2270,1012){\makebox(0,0)[c]{(d)}}

    \put(1500,2200){\makebox(0,0)[c]{Cu}}
    \put(3420,2200){\makebox(0,0)[c]{Cu}}
    \put(1500,1012){\makebox(0,0)[c]{Ag}}
    \put(3420,1012){\makebox(0,0)[c]{Ag}}

  \end{picture}%

  \endgroup
  \end{center}
  \vspace{1.0cm}
  \caption{Excitation spectra, $\nex$, for hydrogen (solid red lines)
    and deuterium (dashed green lines) atoms upon reaching the turning point
    above the copper, panels (a) and (b), and silver, (c) and (d), surfaces.
    Panels (a) and (c) show spectra on linear scales, while a logarithmic
    scale for the excitation spectrum is used in panels (b) and (d) together
    with a larger energy range.}   
    \label{fig:spec}
\end{figure}

The spectrum of electronic excitations generated by the approach of the
adsorbing hydrogen atom is plotted in Fig. \ref{fig:spec} for each of the four
systems under consideration. 
The effects due to the majority and minority spins have been summed in these
spectra. 
Each spectrum consists of a pair of sharp peaks close to the Fermi level, with
electrons being excited just above $\ep_F$ and holes just below.
Away from the Fermi level, i.e. $\vert\ep-\ep_F\vert \geq 0.2$ eV, the
excitation spectra falls off roughly exponentially with energy for both
electrons and holes. 
The shape of the different sections of the excitation spectra, their
dependence on the parameters of the system and the impact of temperature
have been explored previously \cite{mizielinski07,thesis}. 

The isotope effect for the two systems, i.e. the difference between the
excitation spectra for hydrogen and deuterium atoms, appears to be small on
the linear scales used in panels (a) and (b) of Fig. \ref{fig:spec}.
However, the semi-logarithmic scales used to display the same data in panels
(c) and (d) show that the probability of high-energy excitations falls
roughly exponentially, with different rates for the two isotopes.
It has become conventional \cite{lindenblatt06nm,
  lindenblatt06ss,krix07,bird08} to describe these distributions using
Boltzmann factors with an effective temperature $\Teff$. 
Values for $\Teff$ have been extracted from each of the spectra shown in
Fig. \ref{fig:spec} and are presented in Table \ref{tab:T_eff}.
These data show that the differences between the two metal surfaces are small,
while the isotope effect is significant.
TDDFT and Newns-Anderson model calculations for the H/Al(111) system
\cite{bird08} yield effective temperatures in the range 1400-1700 K, which
is similar to those obtained here.

\begin{table}
  \caption{Effective temperatures for electrons and holes for
  the spectra plotted in Fig \ref{fig:spec}. Uncertainties in these values
  arising from the least-squares fitting procedure are approximately $\pm$10
  K.} 
  \label{tab:T_eff}
  \begin{center}
  \begin{tabular}{|c|c|c|c|c|}
    \hline
    System & H/Cu & D/Cu & H/Ag & D/Ag \\
    \hline
    $\Teff$ (electrons)
    & 1400 K & 1160 K & 1370 K & 1110 K \\
    $\Teff$ (holes) 
    & 1470 K & 1160 K & 1410 K & 1150 K \\
    \hline
  \end{tabular}
  \end{center}
  \vspace{0.5cm}
\end{table}

The results presented in Fig. \ref{fig:spec} can be used to estimate the
chemicurrents measured in the thin-film Schottky diode experiments of Nienhaus
and coworkers \cite{nienhaus99}. 
The probability of exciting electrons and holes, $\Pchemie$ and $\Pchemih$,
with sufficient energy to be detected in such devices can be estimated by

\begin{eqnarray}
  \Pchemie(\ep > \ep_S) &=& \int^\infty_{\ep_S} d\ep\ \nex(\ep)\
  a(\ep,\ep_S), \\
  \label{eq:prob_e}
  \Pchemih(\ep > \ep_S) &=& \int^\infty_{\ep_S} d\ep\ \vert\nex(-\ep)\vert\
  a(\ep,\ep_S), 
  \label{eq:prob_h}
\end{eqnarray}

where $\ep_S$ is the Schottky barrier height, $\nex(\ep)$ is the total
excitation spectrum and $a(\ep,\ep_S)$ is a geometrical factor, which contains
two components.
The first describes the attenuation of hot electrons or holes as they propagate
through the metal film, while the second describes the probability, given
isotropic emission of the electrons from the adsorption site (within the
metal), that the electron or hole has sufficient normal energy to cross the 
Schottky barrier at the metal-silicon interface. 
The factor $a$ can be expressed as

\begin{eqnarray}
  a(\ep,\ep_S) = \int_0^{\theta_c} d\theta
  \sin(\theta) \exp\left[-\frac{D}{\lambda \cos(\theta)}\right],
  \label{eq:att}
\end{eqnarray}

where $\theta$ is the angle to the surface normal and 
$\theta_c = \cos^{-1}\sqrt{\ep_S/\ep}$ is the angle above which the electron 
or hole will not have enough normal energy to cross the Schottky barrier.
$\lambda$ is the mean free path of electrons (assumed to be independent
of energy) within the metal film, which has a thickness $D$.
In using these expressions a number of additional assumptions are being made:
there is no preference for the direction of propagation of the excitations
within the metal, the metal has a uniform thickness and the Schottky barrier
height is uniform throughout the device.
The probability that an electron with sufficient normal energy is able to
cross the Schottky barrier is also assumed to be unity.

\begin{figure}[p]
  \begin{center}
  \begin{picture}(0,0)%
    \includegraphics{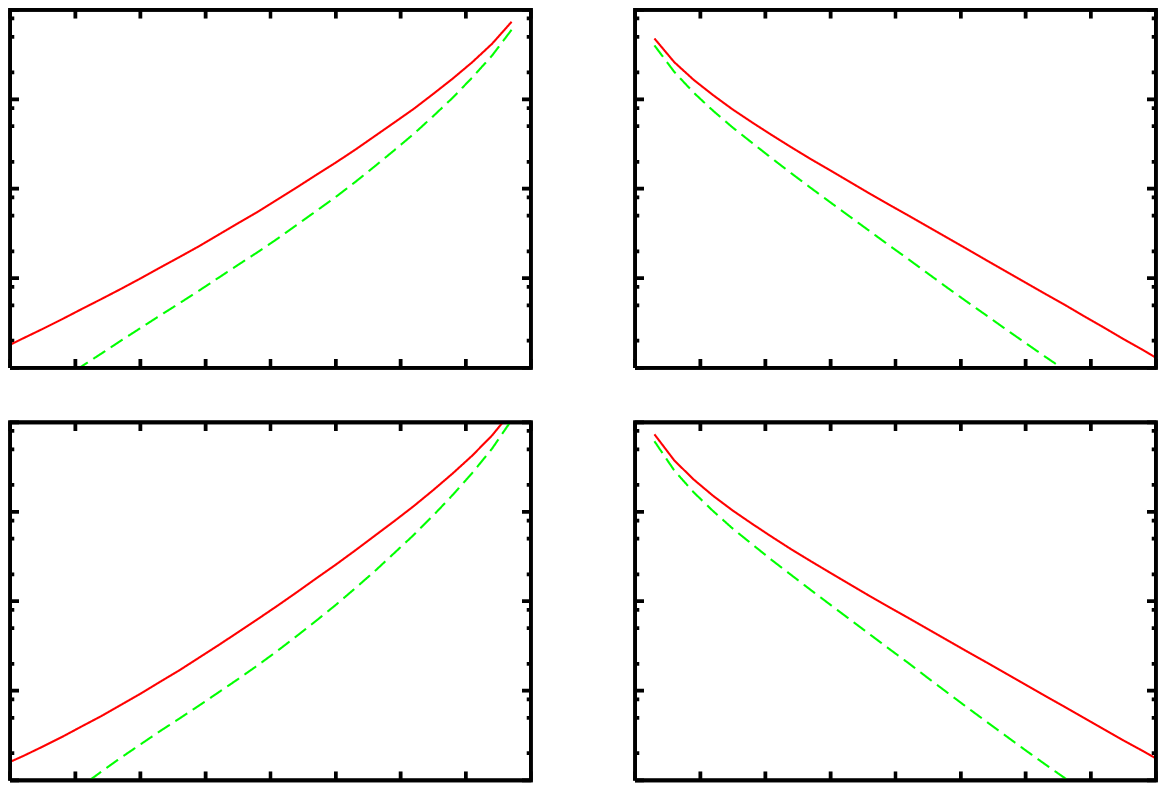}%
  \end{picture}%
  \begingroup
  \setlength{\unitlength}{0.1bp}%
  \begin{picture}(3500,2200)(0,0)%

    \put(3450,   0){\makebox(0,0){ 0.8}}%
    \put(3062,   0){\makebox(0,0){ 0.6}}%
    \put(2675,   0){\makebox(0,0){ 0.4}}%
    \put(2288,   0){\makebox(0,0){ 0.2}}%
    \put(1950,   0){\makebox(0,0){ 0.0}}%
    \put(2675 ,-200){\makebox(0,0)[c]{$\ep_S$ (eV)}}%

    \put(1625,   0){\makebox(0,0)[c]{ 0.0}}%
    \put(1262,   0){\makebox(0,0)[c]{ 0.2}}%
    \put(875 ,   0){\makebox(0,0)[c]{ 0.4}}%
    \put(488 ,   0){\makebox(0,0)[c]{ 0.6}}%
    \put(100 ,   0){\makebox(0,0)[c]{ 0.8}}%
    \put(875 ,-200){\makebox(0,0)[c]{ $\ep_S$ (eV)}}%

    \put(100 ,1131){\makebox(0,0)[r]{ 10$^{-1}$}}%
    \put(100 , 874){\makebox(0,0)[r]{ 10$^{-2}$}}%
    \put(100 , 616){\makebox(0,0)[r]{ 10$^{-3}$}}%
    \put(100 , 358){\makebox(0,0)[r]{ 10$^{-4}$}}%
    \put(100 , 100){\makebox(0,0)[r]{ 10$^{-5}$}}%
    \put(-300, 615){\rotatebox{90}{\makebox(0,0)[c]{$\Pchemi$ (atom$^{-1}$)}}}

    \put(100 ,2319){\makebox(0,0)[r]{ 10$^{-1}$}}%
    \put(100 ,2061){\makebox(0,0)[r]{ 10$^{-2}$}}%
    \put(100 ,1803){\makebox(0,0)[r]{ 10$^{-3}$}}%
    \put(100 ,1545){\makebox(0,0)[r]{ 10$^{-4}$}}%
    \put(100 ,1288){\makebox(0,0)[r]{ 10$^{-5}$}}%
    \put(-300,1803){\rotatebox{90}{\makebox(0,0)[c]{$\Pchemi$ (atom$^{-1}$)}}}
    
    \put(300 ,2200){\makebox(0,0)[c]{(a)}}
    \put(3300,2200){\makebox(0,0)[c]{(b)}}
    \put(300 ,1012){\makebox(0,0)[c]{(c)}}
    \put(3300,1012){\makebox(0,0)[c]{(d)}}

    \put(900 ,2200){\makebox(0,0)[c]{holes}}
    \put(2700,2200){\makebox(0,0)[c]{electrons}}
    \put(900 ,1012){\makebox(0,0)[c]{holes}}
    \put(2700,1012){\makebox(0,0)[c]{electrons}}
  
    \put(2075,1400){\makebox(0,0)[c]{Cu}}
    \put(1500,1400){\makebox(0,0)[c]{Cu}}
    \put(2075, 200){\makebox(0,0)[c]{Ag}}
    \put(1500, 200){\makebox(0,0)[c]{Ag}}
      
  \end{picture}%

  \endgroup
  \end{center}
  \vspace{1.0cm}
  \caption{Probability of measuring electrons and holes in a
    thin-film Schottky device as used by Nienhaus and co-workers. Panels (a)
    and (b) relate to Cu/Si devices, while (c) and (d) refer to Ag/Si
    devices. Electron probabilities are plotted in panels (b) and (d) with
    hole probabilities in (a) and (c). As previously, solid red lines refer to
    calculations for hydrogen atoms and dashed green lines to those for
    deuterium.  A film thickness of 75 \AA{} was assumed with mean-free paths
    of 100 and 240 \AA{} for copper and silver surfaces respectively.}
  \label{fig:chemi_comp}
\end{figure}

The probability of detecting an electron or a hole in a thin-film Schottky
device is plotted in Fig. \ref{fig:chemi_comp}.
Mean-free paths for electrons and holes were taken to be 100 \AA{}
\cite{nienhaus99} for copper and 240 \AA{} \cite{krix07} for silver.
A film thickness of 75 \AA{} was also assumed for both metal films \cite{nienhaus99}.
The magnitude of the chemicurrents measured in the experiments of Nienhaus and
coworkers compares well with these results. 
Experiments using Cu/n-Si(111) devices, with a 0.6 eV Schottky barrier, 
measured $1.5\times10^{-4}$ electrons per incident H-atom \cite{nienhaus99}.
Our model suggests approximately $0.9\times10^{-4}$ electrons per atom for the
single approach to the surface simulated.
Krix, N\"unthel and Nienhaus have recently performed a detailed study of
Ag/p-Si(111) devices which have a well characterised Schottky barrier height
of 0.46 eV \cite{krix07}. 
On exposure to beams of hydrogen and deuterium atoms chemicurrents in the range
1-10$\times10^{-4}$ and 1-$5\times10^{-4}$ holes per atom were measured
respectively.
These data also compare well to our model -- we estimate chemicurrent yields
of $4.8\times10^{-4}$ ($1.5\times10^{-4}$) holes per incident hydrogen
(deuterium) atom. 



\begin{figure}[p]
  \begin{center}
  \begin{picture}(0,0)%
    \includegraphics{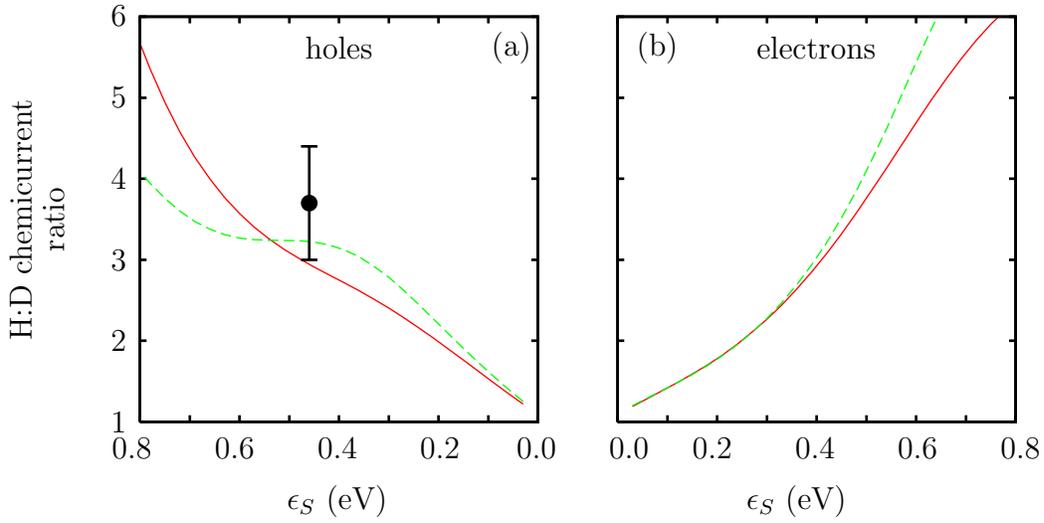}%
  \end{picture}%
  \begingroup
  \setlength{\unitlength}{0.1bp}%
  \begin{picture}(3500,2200)(0,0)%

    \put(3450,   0){\makebox(0,0){ 0.8}}%
    \put(3062,   0){\makebox(0,0){ 0.6}}%
    \put(2675,   0){\makebox(0,0){ 0.4}}%
    \put(2288,   0){\makebox(0,0){ 0.2}}%
    \put(1975,   0){\makebox(0,0){ 0.0}}%
    \put(2625 ,-200){\makebox(0,0)[c]{$\ep_S$ (eV)}}%

    \put(1625,   0){\makebox(0,0)[c]{ 0.0}}%
    \put(1262,   0){\makebox(0,0)[c]{ 0.2}}%
    \put(875 ,   0){\makebox(0,0)[c]{ 0.4}}%
    \put(488 ,   0){\makebox(0,0)[c]{ 0.6}}%
    \put(100 ,   0){\makebox(0,0)[c]{ 0.8}}%
    \put(875 ,-200){\makebox(0,0)[c]{ $\ep_S$ (eV)}}%

    \put(100 ,1631){\makebox(0,0)[r]{ 6}}%
    \put(100 ,1324){\makebox(0,0)[r]{ 5}}%
    \put(100 ,1018){\makebox(0,0)[r]{ 4}}%
    \put(100 , 712){\makebox(0,0)[r]{ 3}}%
    \put(100 , 406){\makebox(0,0)[r]{ 2}}%
    \put(100 , 100){\makebox(0,0)[r]{ 1}}%
    \put(-300, 865){\rotatebox{90}{\makebox(0,0)[c]{H:D chemicurrent}}}
    \put(-175, 865){\rotatebox{90}{\makebox(0,0)[c]{ratio}}}

    \put(1550,1512){\makebox(0,0)[c]{(a)}}
    \put(2100,1512){\makebox(0,0)[c]{(b)}}
  
    \put(900 ,1512){\makebox(0,0)[c]{holes}}
    \put(2700,1512){\makebox(0,0)[c]{electrons}}

  \end{picture}%

  \endgroup
  \end{center}
  \vspace{1.0cm}
  \caption{H:D chemicurrent ratio, $\Pchemi_H/\Pchemi_D$, for the copper
    (solid red lines) and silver (long-dashed green lines) surfaces as a
    function of the Schottky barrier height. Panels (a)
    and (b) refer to hole and electron currents respectively.  The
    point with error bars in (a) is the experimental result reported by Krix
    and co-workers \cite{krix07}.}  
  \label{fig:chemi_ratio}
\end{figure}

There is, however, some uncertainty in experimental measurements of the
absolute chemicurrent yield due to difficulties in quantifying the flux of
atoms reaching the device surface. 
One quantity which is insensitive to this uncertainty is the ratio of the
chemicurrents generated by beams of hydrogen and deuterium atoms.
Krix and co-workers reported that the chemicurrents for hydrogen are
3.7$\pm$0.7 times larger than for deuterium for their Ag/p-Si (111) devices
\cite{krix07}. 
The first experimental report of chemicurrents by Nienhaus and coworkers
\cite{nienhaus99} also estimated the ratio of chemicurrents for Ag/n-Si and
Cu/n-Si devices, giving an electron chemicurrent ratio of approximately six for
barrier heights in the range 0.5-0.6 eV.
To compare our calculations with these results the ratio of chemicurrents
generated by H and D-atoms has been plotted in Fig. \ref{fig:chemi_ratio} for
both copper and silver films. 
For an Ag/p-Si device with a barrier height of 0.46 eV our model yields a
ratio of H:D hole chemicurrents of 3.2:1, while an Ag/n-Si device with 
barrier heights in the range 0.5-0.6 eV gives ratios between 4.1:1 and 5.4:1.
Both results are in good agreement with those reported by Nienhaus, Krix and
co-workers.

\section{Conclusions}
\label{sec:conclusions}

The Newns-Anderson model provides a simple but effective method for analysing
non-adiabatic processes in adsorption at surfaces. It allows for the
calculation of strongly non-adiabatic effects, such as those occuring at a
spin transition. 
Its simplicity means that, for any set of parameter variations, the
calculation of charge and energy transfer rates and the spectrum of electronic
excitations is quick and straightforward.

The key result of this paper is that a single passage of the hydrogen atom
adsorbate through the spin transition yields a significant energy transfer 
into electronic excitations, and sufficient numbers of high energy excitations 
to account for the chemicurrent measured in thin-film Schottky devices. This
leaves open the question of the effect on the chemicurrent yield of the 
remainder of trajectory, as the atom undergoes damped vibrations in the 
surface potential well. It is possible to run the Newns-Anderson model
for such a trajectory, with the damping rate being derived either from
the energy loss in the Newns-Anderson model itself, or using an ab-initio
friction coefficient as in \cite{trail03}. However, a problem arises 
because, at least for the first few oscillations, the hydrogen atom passes 
through the spin transition both when moving towards and moving away from 
the surface. The latter case is where the difficulty emerges.

If we consider the first rebound of the atom within the potential well, 
it can be seen from Fig. \ref{fig:n_E} that the atom has de-polarised
by the time it reaches the back wall and we find that the spin polarisation 
is zero (to numerical accuracy) by the time the atom passes back through 
the altitude where the spin transition occurs. At this point, the atom
can re-polarise, but within our numerical model there is nothing to
determine in which direction the majority spin will be. The re-polarisation
occurs in an unpredictable and irreproducible way, because it is driven
by numerical instability. We do not believe that this can be regarded as 
representing a physical reality. An alternative treatment is to keep the 
atom non-polarised for the remainder of the trajectory after the first
approach to the surface, and to analyse non-adiabaticity by comparing the 
time-dependent solution with a non-polarised adiabatic state (a
metastable, non-polarised adiabatic solution exists even when the ground
state is spin polarised). We have performed such calculations, and the
results show that the full trajectory gives an excitation spectrum whose
magnitude is a factor of about two or three times that shown in Fig.
\ref{fig:spec}. However, the justification for using this treatment is 
not clear; for example, if the atom could escape from the surface the 
method cannot be right because the final state of the atom should be
spin polarised. The correct way to handle re-polarisation of the atom
as it leaves the surface remains an open question, and one that
is as relevant to ab-initio approaches like TDDFT as it is to 
model calculations like the ones presented here.

\section*{Acknowledgments}

Financial support from the Engineering and Physical Sciences Research Council
(EPSRC), Grant No. EP/E021646/1, is gratefully acknowledged.

\end{document}